\documentclass[aps,prb,twocolumn,10pt,superscriptaddress,showpacs,showkeys]{revtex4-1}

\usepackage{amsmath,latexsym,amsfonts,amssymb}
\usepackage{natbib}
\usepackage{epsfig}
\usepackage{graphicx}
\usepackage{inputenc}
\usepackage{setspace}
\usepackage{multirow}
\usepackage{array}
\usepackage{tabularx}
\usepackage{bm}
\usepackage{color}
\usepackage{xcolor}
\usepackage{gensymb}
\usepackage[left]{lineno}
\usepackage{float} 
\usepackage{ulem}
\usepackage{siunitx}

\begin{document}


\title{Structural and magnetic phases of topological kagome metal Fe$_3$Sn$_2$ under pressure}

\author{Sumanta Chattopadhyay}
\email{sumanta@csr.res.in}
\affiliation{UGC-DAE Consortium for Scientific Research Mumbai Centre, 246-C CFB, BARC Campus, Mumbai 400085, India}

\author{Laure Thomarat}
\affiliation{Université Paris-Saclay, CNRS, Laboratoire de Physique des Solides, 91405, Orsay, France.}

\author{Chin Shen Ong}
\affiliation{Department of Physics and Astronomy, Uppsala University, 751 20, Uppsala, Sweden}

\author{Kuldeep Kargeti}
\affiliation{Department of Physics and Astronomy, Uppsala University, 751 20, Uppsala, Sweden}
\affiliation{Department of Physics, Bennett University, Greater Noida, Uttar Pradesh 201310, India}

\author{Lipika}
\affiliation{Indian Institute of Technology-Delhi, New Delhi, Hauz Khas, 110016, India}

\author{Jean-Pascal Rueff}
\author{Lucie Nataf}
\affiliation{Synchrotron SOLEIL, L'Orme des Merisiers, Saint Aubin BP 48, 91192, Gif-sur-Yvette, France}

\author{Kaustuv Manna}
\affiliation{Indian Institute of Technology-Delhi, New Delhi, Hauz Khas, 110016, India}

\author{Swarup K. Panda}
\affiliation{Department of Physics, Bennett University, Greater Noida, Uttar Pradesh 201310, India}

\author{Chandra Shekhar}
\affiliation{Max Planck Institute for Chemical Physics of Solids, 01187 Dresden, Germany}

\author{Victor Balédent}
\email{victor.baledent@universite-paris-saclay.fr}
\affiliation{Université Paris-Saclay, CNRS, Laboratoire de Physique des Solides, 91405, Orsay, France.}
\affiliation{Institut universitaire de France (IUF)}

\begin{abstract}
We investigate the pressure-induced evolution of crystal structure and magnetism in the kagome ferromagnet Fe$_3$Sn$_2$ by combining X-ray diffraction, X-ray Emission Spectroscopy, X-ray Magnetic Circular Dichroism, and spin-polarized density functional theory calculations. X-ray diffraction reveals a structural phase transition above $\sim$20~GPa, which coincides with a pronounced reduction of the local Fe magnetic moment evidenced by X-ray emission spectroscopy, indicating a high-spin to low-spin transition. While XES probes the amplitude of the local moment, XMCD provides direct information on the orientation of the ordered magnetic moments and uncovers a rich pressure--temperature magnetic phase diagram. At room temperature, a collinear ferromagnetic phase with moments aligned along the $c$ axis persists up to the structural transition. At low temperature, a tilted magnetic configuration remains stable to significantly higher pressures, while at intermediate temperatures pressure stabilizes the low-temperature magnetic phase at the expense of the high-temperature one. Spin-polarized first-principles calculations show that, although isotropic ferromagnetic exchange interactions remain robust under compression, pressure enhances spin--orbit--driven magnetic anisotropy and Dzyaloshinskii--Moriya interactions, favoring non-collinear magnetic configurations. Our results demonstrate that pressure reshapes the magnetic energy landscape of Fe$_3$Sn$_2$ by coupling lattice, spin state, and relativistic magnetic interactions, establishing hydrostatic pressure as an effective control parameter to engineer magnetic anisotropy and potentially topological phases in kagome materials.
\end{abstract}


\maketitle{}

\section{Introduction}
Magnetic structure and its evolution under external perturbations are central topics in the study of quantum materials. In particular, kagome lattices, with their geometry of corner-sharing triangles, provide a unique platform in which magnetic interactions and electronic topology are intertwined. While magnetic frustration was initially the main focus in kagome systems, leading to proposals of exotic ground states such as quantum spin liquids \cite{Balents2010, Zhou2017}, it has become clear that kagome materials also host topologically non-trivial band structures with flat bands and Dirac cones \cite{Ye2018, Yin2019}. The orientation of the magnetic moments, and more generally the magnetic structure, plays a crucial role in determining the electronic topology, including the appearance of Weyl nodes, Chern gaps, and anomalous Hall effects \cite{Wang2016, Liu2018}. Therefore, controlling and understanding changes in magnetic orientation provide a powerful route to engineer emergent topological and correlated phases.

Among kagome compounds, Fe-Sn based materials stand out as key systems hosting this interplay of magnetism and topology. FeSn, with its simple crystal structure, has been widely investigated as a model kagome magnet \cite{Sales2019, Kang2020, Li2022, Tao2023}. In contrast, $Fe_3Sn_2$ exhibits a more complex bilayer kagome structure and richer physics. It is a soft ferromagnet with a high Curie temperature ($T_C = 657$ K) \cite{Malaman1978}, and exhibits a giant anomalous Hall effect arising from Berry curvature associated with Dirac-like bands \cite{Wang2016}. Recent studies have shown that the spin orientation in $Fe_3Sn_2$ strongly affects its electronic structure \cite{Fenner2009}, even tuning the system into Weyl semimetallic phases \cite{Ren2022}. Furthermore, real-space topological textures such as skyrmions have been observed \cite{Zhipeng2017}, highlighting the coexistence of reciprocal and real-space topology. Beyond the magnitude of the magnetic moment, spin-orbit coupling and magnetocrystalline anisotropy play a central role in kagome ferromagnets, as they control the orientation of the magnetic moments and enable non-trivial electronic topology. In $Fe_3Sn_2$, subtle changes in crystal symmetry or lattice parameters can therefore strongly modify the balance between collinear and non-collinear magnetic configurations.

A key challenge is to explore how the magnetic orientation and amplitude respond to external parameters. Magnetic field has already been shown to induce spin reorientations in $Fe_3Sn_2$ thin films \cite{Ren2022}. Strain has been theoretically proposed as an alternative tuning parameter in kagome systems, capable of acting as an effective magnetic field and modifying exchange interactions and lattice symmetry \cite{Liu2020}. Among the different external stimuli, hydrostatic pressure is particularly powerful because it directly modifies interatomic distances, bandwidths, and exchange pathways, potentially coupling structural and magnetic degrees of freedom \cite{Wang2012}. Pressure can induce structural phase transitions that dramatically reshape the magnetic structure or even quench the magnetic moment \cite{Kunes2010}. Despite its importance, the pressure dependence of magnetism in kagome magnets remains largely unexplored.

In this context, $Fe_3Sn_2$ constitutes an ideal platform to investigate how magnetic structure, in particular spin orientation and magnetic moment amplitude, evolve under pressure. Its layered kagome-based structure, intrinsic ferromagnetism, and strong spin-orbit coupling make it highly sensitive to lattice distortions. Moreover, the close interplay between crystal symmetry and electronic topology suggests that pressure could unlock new magnetic or topological phases. Previous works have hinted at a possible structural transition at high pressure \cite{Giefers2006}, but the magnetic consequences of such a transition have not been addressed experimentally.

In this paper, we combine X-ray diffraction, X-ray Magnetic Circular Dichroism, X-ray Emission Spectroscopy, and density functional theory calculations to systematically investigate the pressure evolution of structure and magnetism in $Fe_3Sn_2$. We focus on the modification of magnetic orientation and the collapse of ferromagnetism across the pressure-induced structural transition. We demonstrate that pressure stabilizes distinct magnetic regimes and induces a transformation of the magnetic state intimately linked to structural changes. By disentangling the evolution of the local magnetic moment amplitude from that of magnetic anisotropy and spin orientation, we provide a comprehensive picture of how pressure reshapes the magnetic energy landscape of this kagome compound. Combining spectroscopy, diffraction, and spin-polarized first-principles calculations, we show that pressure not only drives a spin-state transition but also enhances spin-orbit-driven anisotropic interactions, stabilizing distinct magnetic configurations over extended pressure--temperature ranges.

\section{X-Ray Diffraction}
\begin{figure}[htbp]
\includegraphics[width=0.95\linewidth, angle=0]{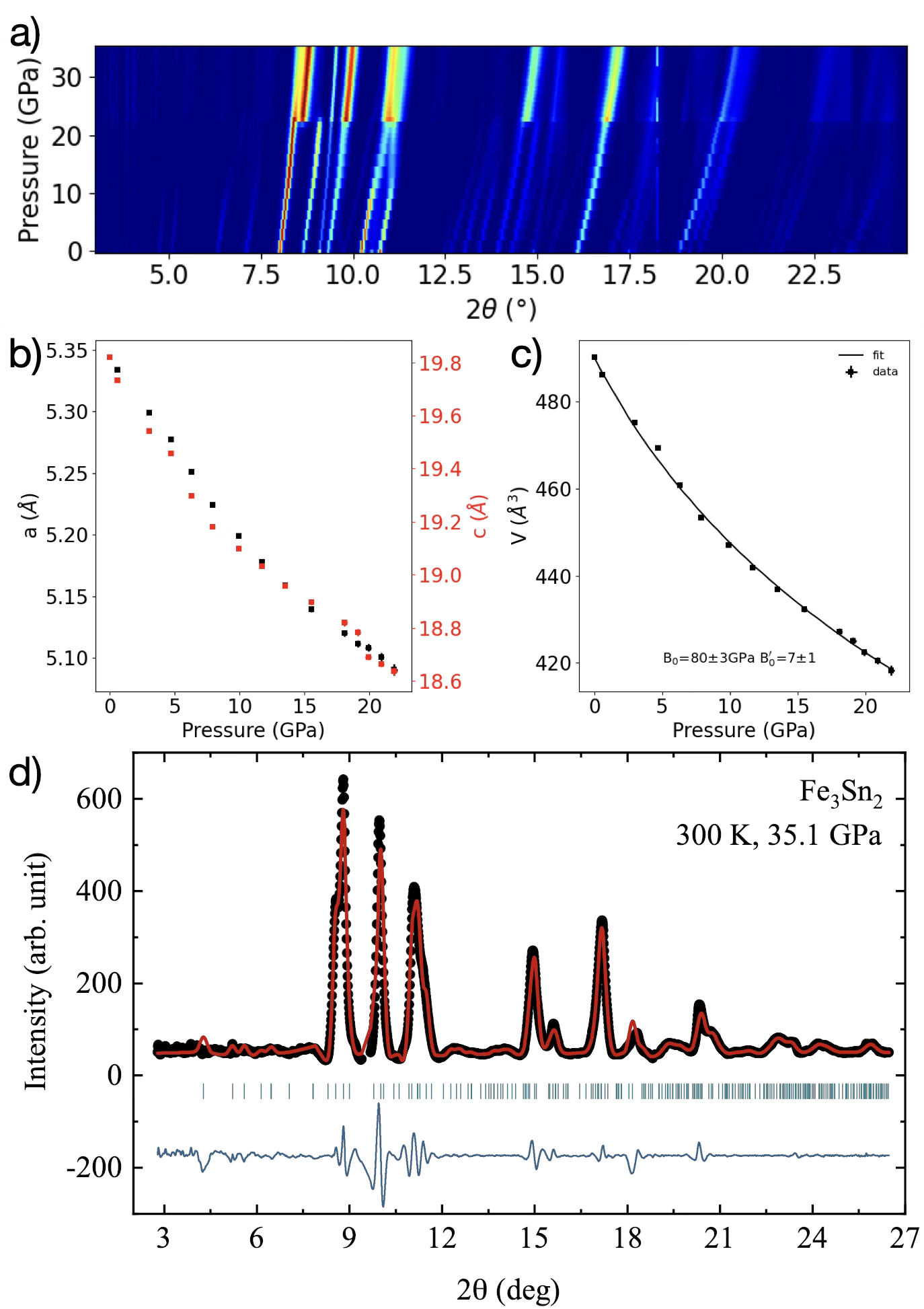}
\caption{(Color online) High-pressure X-ray diffraction analysis. (a) Color map showing the pressure evolution of X-ray diffraction patterns, revealing the emergence of additional peaks above 22 GPa, indicative of a structural phase transition. (b) Pressure dependence of the refined lattice parameters a (black symbols) and c (red symbols) obtained from Lebail refinement using the $R\bar{3}m$ space group up to the transition pressure. (c) Unit cell volume as a function of pressure (black squares) fitted with a third-order Birch–Murnaghan equation of state (solid black line). (d) Experimental high pressure (35.1 GPa) diffractogramm (black points) together with Le Bail fit (red line) using only the high pressure space group $Pnma$. The blue line is the difference between experimental data and fitted curve.} 
\label{XRD}
\end{figure}

The high-resolution x-ray diffraction (XRD) was carried out at the PSICHE beamline of the SOLEIL synchrotron with a photon energy of 23 keV ($\lambda$ = 0.3738 Å). $Fe_3Sn_2$ powder sample was mounted in a Diamond Anvil Cell with 300$\mu$m culet diamonds, equiped with Rhenium gasket and ethanol:methanol mixture in 1:4 volumic proportion as pressure transmitting medium \cite{Tateiwa2009}. Gold powder was also put in the chamber to monitor the pressure, using and the measured lattice parameter and the Au equation of state \cite{Boettger2012}. Measured were performed at room temperature (300 K) between 0 and 35 GPa. The results are presented in the colormap of Fig. \ref{XRD}a. Beyond the expected contraction of the lattice visible by the continuous shift of the peaks to high angles, we observe a structural transition around 22 GPa confirming what was observed in a previous publication \cite{Giefers2006}. This transition corresponds to a symmetry lowering toward a lower-symmetry high-pressure phase, whose average crystallographic symmetry will be discussed in connection with DFT calculations.

We represent in Fig. \ref{XRD}b and c the extracted lattice parameters using the Lebail method, and the evolution of the unit cell volume fitted by a Birch–Murnaghan equation \cite{Birch1947}. The results are similar to the one obtained previously, with $K_0$ = 79 $\pm$ 5 GPa and $K_0'$ 7$\pm$1 for the Bulk modulus and its derivative respectively. 

\section{Density functional theory}
In order to further characterize the high pressure structure, we carried out a structural search using the evolutionary algorithm implemented in the \texttt{USPEX} package~\cite{Oganov2006,Oganov2011,Lyakhov2013}, in variable-composition mode. The population size was set to 60, with an initial pool of 100 randomly generated structures, and evolution continued for up to 10 generations. Structural candidates were ranked based on their enthalpy, and the selection and variation of structures were governed by heredity (50\%), random symmetric structures (20\%), soft mutation (20\%), lattice mutation (10\%), and permutation (10\%). Structures containing between 15 and 30 atoms were allowed, and the search was unconstrained with respect to the crystallographic space groups. The \texttt{USPEX} code interfaced with \texttt{Quantum Espresso}, using scalar relativistic pseudopotentials from PSlibrary. The local density approximation (LDA) was used for the electron exchange and correlation energy, following the projector augmented wave (PAW) formalism of Kresse and Joubert~\cite{Kresse1999}. The plane-wave cutoff for the DFT calculation was set to 50~Ry for the plane-wave expansion of the wavefunctions.

For each generation of structures, DFT calculations were conducted over four sequential calculations, each employing increasingly stringent convergence criteria and relaxation settings.  The convergence thresholds and electronic smearing widths were progressively tightened across these stages to ensure numerical stability and reliable identification of the lowest-enthalpy candidates. In the last calculation, the Brillouin zone was sampled using Monkhorst–Pack grids generated to a reciprocal-space resolution of 0.5~\AA$^{-1}$. The self-consistent DFT calculation was considered converged when the total energy difference between successive iterations dropped below $10^{-6}$~Ry. Ionic relaxations were performed using the Broyden–Fletcher–Goldfarb–Shanno (BFGS) quasi-Newton algorithm, until the forces on all atoms were smaller than $10^{-3}$~Ry/Bohr and the residual pressure deviated by less than 0.5~kbar from the target external pressure of 45~GPa. This pressure, which exceeds the 35~GPa used in the diffractogram, was chosen to ensure that the system had fully undergone the intended phase transition.

In our structural search using \texttt{USPEX}, we identified a low-enthalpy structure with orthorhombic symmetry, belonging to the $Pnma$ space group. To assess its relevance to experiment, we used it as the basis for X-ray diffraction (XRD) analysis. A Le Bail fit of the experimental XRD data was performed using only the space group symmetry ($Pnma$) and the optimized lattice parameters ($a=6.6283$~Å, $b=4.8723$~Å, $c=7.6609$~Å), yielding excellent agreement with the measured XRD spectrum (Fig.~\ref{XRD} d ). The high-pressure $Pnma$ phase represents a clear departure from the ambient-pressure $R\bar{3}m$ phase, which fails to reproduce key features of the observed diffraction pattern. This supports a pressure-induced structural transition, which we will further correlate with other spectroscopic signatures in XMCD and XAS measurements. The fact that the atomic positions of the predicted structure do not fully reproduce the experimental diffraction pattern suggests that the real high-pressure phase may involve additional degrees of freedom, such as superstructural modulations, disorder, or fractional occupancies\cite{Ong2024}, which are not captured in the present model.

\section{X-ray Emission Spectroscopy}
\begin{figure}[htbp]
\includegraphics[width=0.9\linewidth, angle=0]{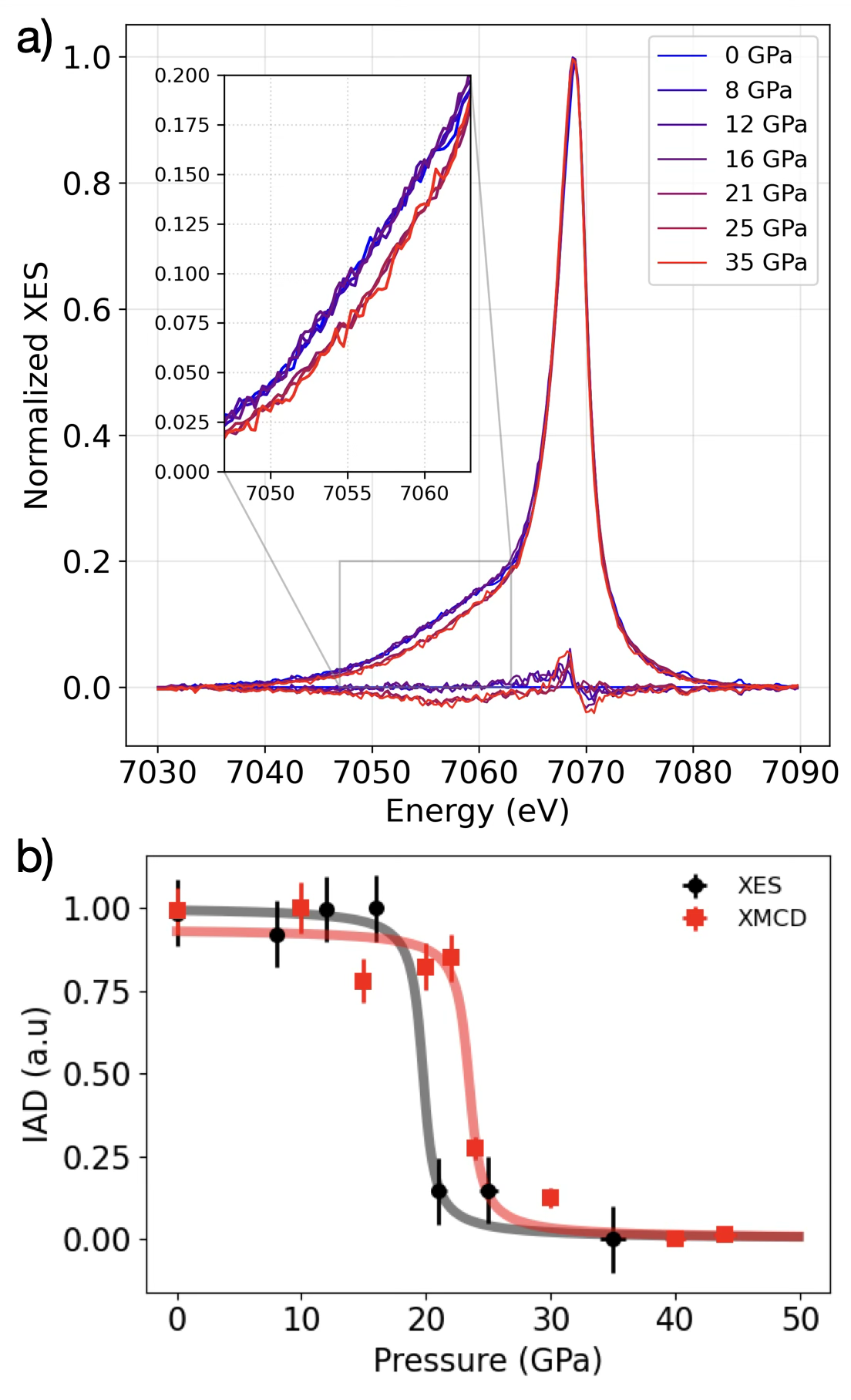}
\caption{(Color online) (a) Fe K$_\beta$ x-ray emission in $Fe_3Sn_2$ as a function of pressure measured at 300 K. The difference with 0 GPa is also plotted. (b) Evolution of the Integrated Amplitude Difference of the XES spectra as function of pressure (in black) compared to the evolution of the dichroic signal associated to ferromagnetism (in red). }
\label{XES}
\end{figure}

To probe the evolution of the local magnetic moment under pressure, we performed X-ray Emission Spectroscopy at the RIXS endstation of the GALAXIES beamline at synchrotron SOLEIL \cite{Rueff2015, Ablett2025}. Pressure was applied using a membrane-driven diamond anvil cell (DAC) equipped with 1.2 mm thick diamonds featuring 300 $\mu$m culets. Multiple $Fe_3Sn_2$ single crystals were placed in a 150 $\mu$m hole within a stainless steel gasket, along with ruby chips for in situ pressure measurement \cite{Shen2020}, and Daphne oil as the pressure-transmitting medium \cite{Tateiwa2009}. X-ray emission spectroscopy (XES) were performed at room temperature (300 K) in transmission geometry, employing a 1-meter-radius spherically bent Ge(620) crystal analyzer and an avalanche photodiode detector arranged in the Rowland circle geometry. The total energy resolution at the Fe K$_\beta$ line (~7057 eV, $\theta_B$=79\degree) was 1.1 eV (full width at half maximum, FWHM). The XES spectra were collected using incident x-rays at 10 keV. The emission spectra are presented in Fig. \ref{XES}a. The K$_{\beta'}$ satellite intensity reflects the 3p--3d exchange interaction and is directly sensitive to the amplitude of the local magnetic moment. 
The K$_\beta'$ satellite intensity reflects the 3p--3d exchange interaction and is directly sensitive to the amplitude of the local magnetic moment. We observe two pressure ranges for which all the spectra overlap: below (blue curves) and above (red curves) $\sim$20 GPa. The reduction in spectral weight at the low-energy shoulder indicates a decrease in the amplitude of the local magnetic moment of iron.
This amplitude can be plotted as a function of pressure using the integrated absolute difference (IAD) method, as shown in Fig.~\ref{XES}b (black circles). A clear high-spin to low-spin transition is observed around 20 GPa. This transition coincides with the pressure-induced structural transition identified by X-ray diffraction and supported by DFT calculations, demonstrating a strong coupling between lattice degrees of freedom and the local spin state. The slight shift in the transition pressure between the XES and XMCD measurements, of approximately 2 GPa, can be attributed to the different pressure-transmitting media used (Daphne oil for XES and an ethanol:methanol mixture for XMCD, respectively). Such a coupling is expected in kagome systems where changes in Fe–Fe and Fe–Sn bonding strongly affect bandwidths and exchange interactions. Nonetheless, XES only provides access to the amplitude of the local moment and does not contain information about its orientation. To address this point, we turn to X-ray Magnetic Circular Dichroism.

\section{X-ray Magnetic Circular Dichroism}
\begin{figure}[htbp]
\includegraphics[width=0.99\linewidth, angle=0]{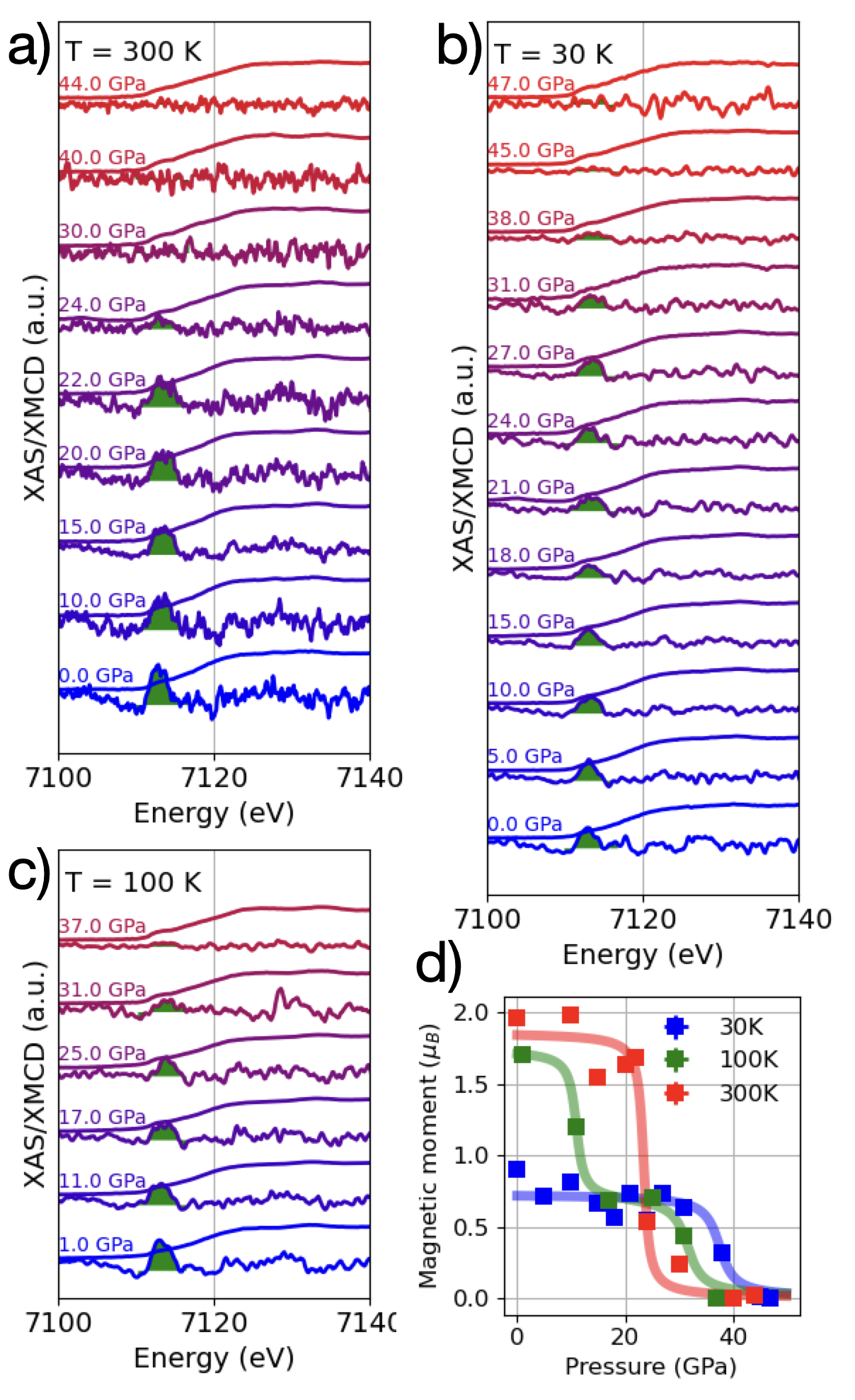}
\caption{(Color online) (a-c) : X-ray absorption spectra and associated dichroic signal at Fe K-edge, performed at 1.3 T and 300, 30 and 100 K respectively. Dichroic signals are multiplied by a factor of 500 for a sake of visibility. The green area correspond to the intensity reported in panel (d). (d) Evolution of the ferromagnetic component along the $c$ axis deduced from the dichroic area (in green in (a-c)) and normalized to 2 $\mu_B$ at ambient condition (300 K and 0 GPa), for 30 K (blue), 100 K (green) and 300 K (red).}
\label{XMCD}
\end{figure}

X-ray Magnetic Circular Dichroism (XMCD) is sensitive to the projection of the magnetic moment along the X-ray propagation direction and therefore provides direct information on the magnetic orientation. In this sense, XMCD is complementary to XES and DFT: while the latter establish the existence and amplitude of the local magnetic moment, XMCD resolves how this moment is oriented with respect to the crystal axes and how magnetic anisotropy evolves under pressure. XMCD measurements on $Fe_3Sn_2$ single crystals were performed at the ODE beamline of the SOLEIL synchrotron, focusing on the Fe K-edge \cite{Baudelet2011, Baudelet2016}. Pressure was applied using a diamond anvil cell (DAC) equipped with diamonds featuring 300 µm culet size. The $Fe_3Sn_2$ crystals were loaded into a rhenium gasket with a 150 µm diameter and 40 µm thick hole, along with a ruby chip for pressure monitoring via standard ruby fluorescence calibration \cite{Shen2020}. Due to the crystal shape, there was a preferred orientation inside the chamber, with the c-axis positioned perpendicular to the diamond faces. An ethanol:methanol mixture in a 1:4 volumetric ratio was employed as the pressure-transmitting medium \cite{Tateiwa2009}. Data were collected in three separate experiments at constant temperatures of 30 K, 100 K, and room temperature (300 K), with pressure progressively increased. The XMCD signal was measured by reversing a 1.3 T magnetic field, with approximately a dozen cycles performed for each pressure point. The absorption spectra were normalized to unity at high energy, and the dichroic signal was similarly normalized. X-ray absorption spectroscopy (XAS) spectra, alongside the XMCD signal (magnified by a factor of 500 for clarity), are shown in Figs. \ref{XMCD}a-c.

\textbf{300 K – High-temperature magnetic phase.}  
At ambient temperature, $Fe_3Sn_2$ is known to have a ferromagnetic structure with moments aligned nearly parallel to the $c$ axis. Our XMCD data confirm that the dichroic signal remains strong and positive up to $\sim$20 GPa, indicating that the magnetic moments retain their orientation along the $c$ axis in this pressure range. At around 20 GPa, the XMCD intensity collapses sharply. This collapse coincides with both the spin-state transition observed by XES and the structural transition revealed by XRD, indicating that at 300 K a single magnetic phase with $M \parallel c$ is stable up to the pressure-induced structural instability.

\textbf{30 K – Low-temperature tilted phase.}  
At 30 K and ambient pressure, $Fe_3Sn_2$ adopts a magnetic configuration in which the spins are tilted away from the $c$ axis. In our XMCD geometry, this results in a reduced dichroic signal. Remarkably, this signal persists up to at least 33--39 GPa. The significantly higher transition pressure at low temperature suggests that the low-temperature magnetic phase is more robust against compression, consistent with the general increase of structural transition pressures at low temperature. This robustness may also be related to a pressure-induced modification of magnetic anisotropy, expected when the crystal symmetry is reduced, as suggested by the DFT results.

\textbf{100 K – Intermediate regime.}  
At 100 K, an intermediate behavior is observed. At low pressure, the XMCD signal resembles that at 300 K, indicating moments aligned close to the $c$ axis. However, around 10 GPa, the XMCD amplitude decreases and stabilizes at a value comparable to that of the low-temperature tilted phase. This demonstrates that under pressure the low-temperature magnetic configuration becomes stabilized at significantly higher temperatures than at ambient pressure. This phase remains stable up to $\sim$30 GPa.

\textbf{Pressure–temperature magnetic phase diagram.}  
Altogether, XMCD reveals that pressure destabilizes the high-temperature collinear magnetic phase in favor of the low-temperature tilted configuration. Under compression, the phase boundary between these two magnetic states shifts to higher temperatures, indicating that pressure reshapes the magnetic anisotropy landscape. This behavior highlights a strong spin–lattice coupling in $Fe_3Sn_2$, reminiscent of other iron-based kagome compounds such as $Fe_3Ge$ \cite{Lou2024}. 

\section{Spin-polarized DFT}
To gain microscopic understanding of the observed pressure-induced evolution of the magnetic behaviors in Fe$_3$Sn$_2$, we performed spin-polarized density functional theory (DFT) calculations within the plane-wave pseudo-potential formalism, as implemented in the \textsc{VASP} package~\cite{vasp1,vasp2}. The exchange–correlation effects were treated using the generalized gradient approximation (GGA), and the on-site Coulomb interaction on Fe-$3d$ orbitals was incorporated through the GGA+$U$ approach to account for electronic correlation effects beyond conventional DFT. Two representative values of the Hubbard parameter, $U=2$~eV and $U=4$~eV, were considered to assess the influence of electronic correlations on structural and magnetic properties. The experimentally determined rhombohedral crystal structure (space group $R\bar{3}m$) was used as the starting geometry, with lattice parameters $a=b=5.34$~\AA{} and $c=19.80$~\AA{} at ambient pressure~\cite{Giefers2006}. To simulate the effect of hydrostatic pressure, we uniformly scaled the lattice parameters while maintaining a constant $c/a$ ratio in order to preserve the symmetry. The unit cell volume ($V$) was systematically varied from $-5\%$ ($\frac{V-V_0}{V_0}$) to $0\%$ relative to the ambient experimental value ($V_0$). For each volume, all internal atomic coordinates were fully relaxed until the residual Hellmann–Feynman forces on every atom were smaller than $10^{-3}$~eV/\AA. 
This computational strategy enables us to track the evolution of electronic structure and magnetic order as a function of pressure, thereby providing theoretical insight complementary to the experimental observations.
\par 
To understand the microscopic magnetic interactions in Fe$_3$Sn$_2$ and their evolution under pressure, we computed all relevant terms of the effective spin Hamiltonian—namely the isotropic Heisenberg exchange, the antisymmetric Dzyaloshinskii–Moriya (DM) interaction, and the symmetric anisotropic exchange. These quantities were evaluated at both the ambient equilibrium volume and under $5\%$ hydrostatic compression to assess the pressure dependence of magnetic couplings. To capture spin–orbit–driven anisotropic interactions, the  Lichtenstein–Katsnelson–Antropov–Gubanov (LKAG) methodology was extended to its relativistic form using the Korringa–Kohn–Rostoker (KKR) Green-function framework implemented in \textsc{RSPt} code. This method has been discussed in details in Ref.~\cite{kargeti2025charge,kargeti2024strain} and shown to provide a reliable description of magnetic interactions in transition metal compounds. This yields the full exchange tensor $J_{ij}^{\alpha\beta}$, from which the isotropic exchange, DM interaction vectors, and symmetric anisotropic exchange components are extracted. The single-ion anisotropy (SIA) was computed independently from fully relativistic total-energy calculations with the magnetic quantization axis constrained along different crystallographic directions. 

\begin{table}[t]
\caption{Equilibrium lattice parameters obtained from GGA and GGA+$U$ calculations compared with experimental results. The experimental bulk modulus ($B$) is not available (NA).}
\label{lattice_parameters}
\centering
\begin{tabular}{lcccc}
\hline\hline
\textbf{Method} & \textbf{$a$ (Å)} & \textbf{$c$ (Å)} & \textbf{$V$ (Å$^3$)} & \textbf{$B$ (GPa)} \\
\hline
GGA                 & 5.33 & 19.78 & 488.26 & 102.8 \\
GGA+$U$ ($U = 2$ eV) & 5.38 & 19.97 & 502.09 &  81.6 \\
GGA+$U$ ($U = 4$ eV) & 5.45 & 20.21 & 520.42 &  80.1 \\
Experiment          & 5.34 & 19.80 & 488.89 &  NA   \\
\hline\hline
\end{tabular}
\end{table}

\begin{figure*}
    \begin{center}
     \includegraphics[width=1.7\columnwidth]{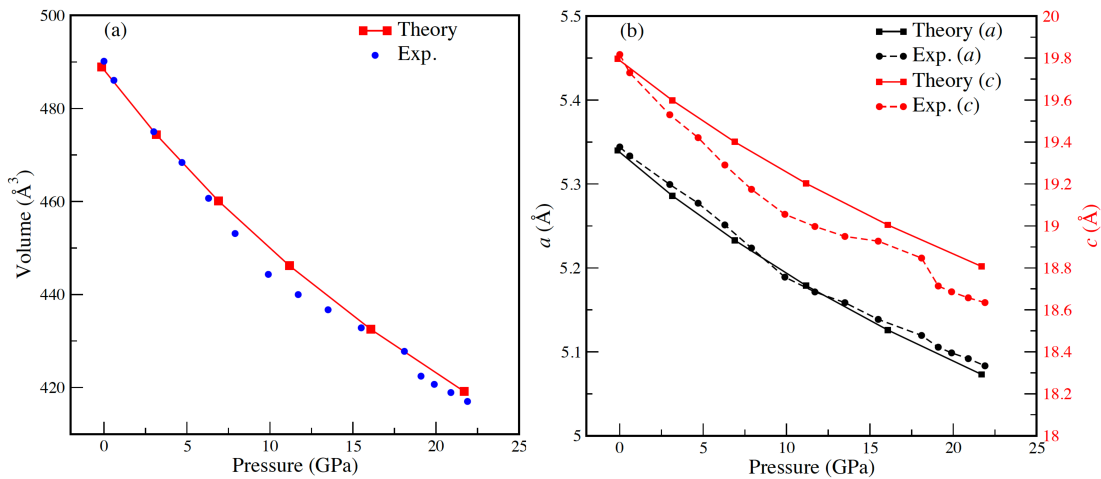}
    \caption{Comparison between experimental and theoretical structural parameters of $Fe_3Sn_2$ under pressure. (a) Pressure dependence of the unit-cell volume. (b) Pressure dependence of the lattice parameters. Experimental data are compared with results obtained from density functional theory calculations, highlighting the agreement across the pressure-induced structural transition.}
    \label{volume}    
    \end{center}
\end{figure*}

\subsection{Optimized lattice parameters}
We first examined the magnetic stability by computing the total energies of two representative spin configurations, namely ferromagnetic (FM) and antiferromagnetic (AFM) states. The results obtained from both GGA and GGA+$U$ ($U=2$ and $4$~eV) formalisms show FM configuration remained energetically favored. This theoretical result is consistent with experimental XMCD measurements, which indicate a predominantly ferromagnetic character in ambient condition. Further, we computed total energies as a function of volume for both GGA and GGA+$U$ ($U=2$~eV and $4$~eV) approaches in the ferromagnetic state. The resulting energy–volume data were fitted using the third-order Birch–Murnaghan equation of state~\cite{Birch,murnaghan}, from which equilibrium lattice parameters and bulk moduli were obtained as shown in Table~\ref{lattice_parameters}). The equilibrium lattice parameters derived from GGA calculations show excellent agreement with experimental values, while the use of Hubbard $U$ leads to a moderate lattice expansion. Consequently, the GGA approximation was chosen for subsequent pressure-dependent analyses. 
\par 
In addition to the energy–volume analysis, the pressure corresponding to each compressed volume was obtained directly from the derivative of the third-order Birch–Murnaghan equation of state. This allowed us to construct the theoretical pressure–volume ($p$–$V$) curve and compare it quantitatively with the available experimental data. As shown in Fig.~\ref{volume}(a), the $p$–$V$ relation calculated within the GGA framework reproduces the experimental trend remarkably well over the entire pressure range studied, indicating that the chosen exchange–correlation treatment captures the underlying compressibility and elastic response of Fe$_3$Sn$_2$ with high fidelity. Furthermore, the individual lattice parameters $a(p)$ and $c(p)$, extracted at each volume, exhibit pressure dependence consistent with experimental observations (see Fig.~\ref{volume}(b)). The close agreement between theory and experiment reinforces the reliability of the structural description obtained within GGA. The combination of GGA-based total-energy calculations and equation-of-state analysis thus provides a solid microscopic foundation for exploring the pressure-induced evolution of structural, electronic, and magnetic degrees of freedom in Fe$_3$Sn$_2$.

\begin{figure}
    \begin{center}
     \includegraphics[width=0.8\columnwidth]{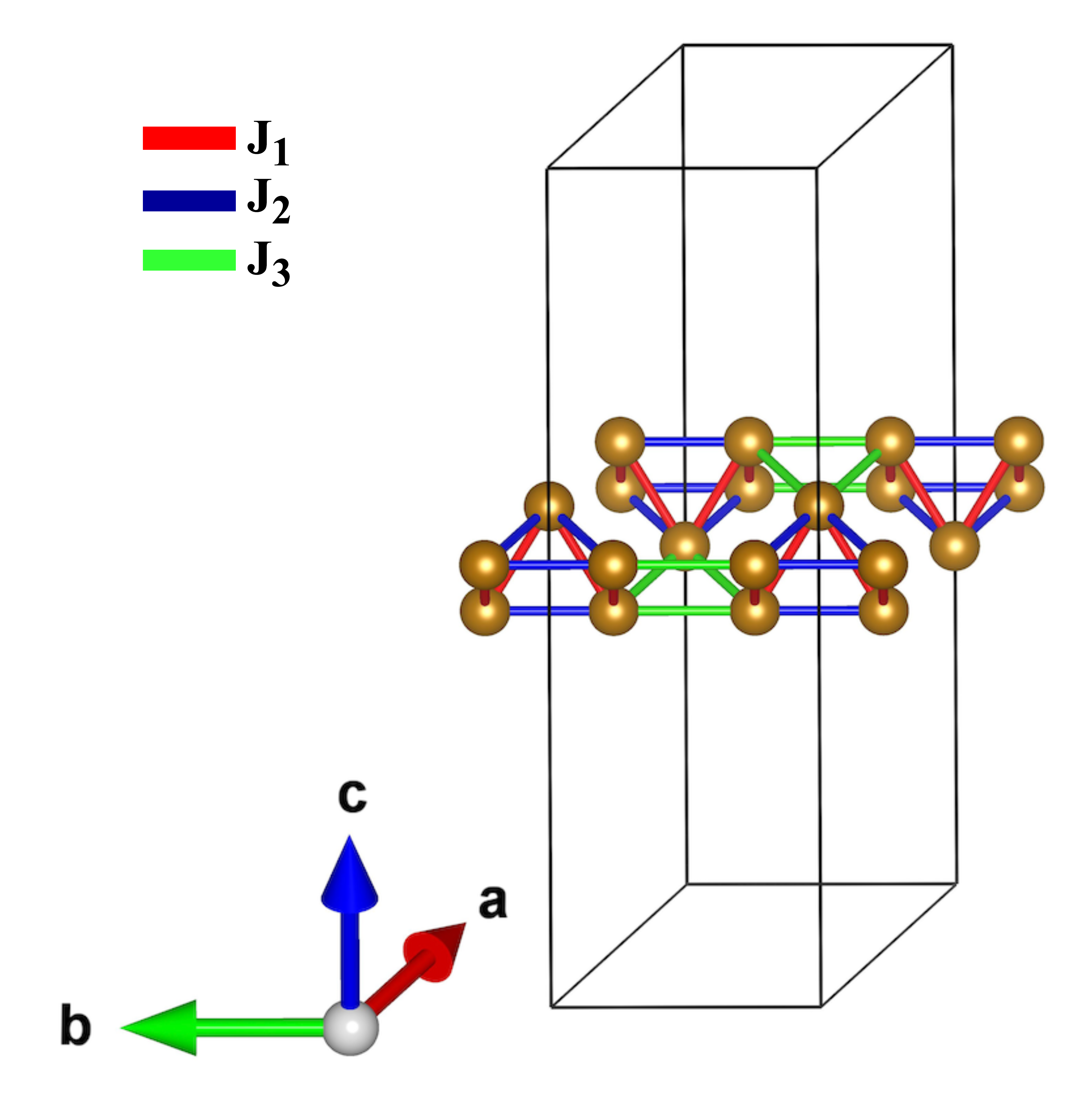}
    \caption {Colour representation of dominant exchange interactions in the unit cell of Fe$_3$Sn$_2$.}
    \label{exchange}    
    \end{center}
\end{figure}

\subsection{Magnetism under application of pressure}
\begin{figure}[t]
    \begin{center}
     \includegraphics[width=1.0\columnwidth]{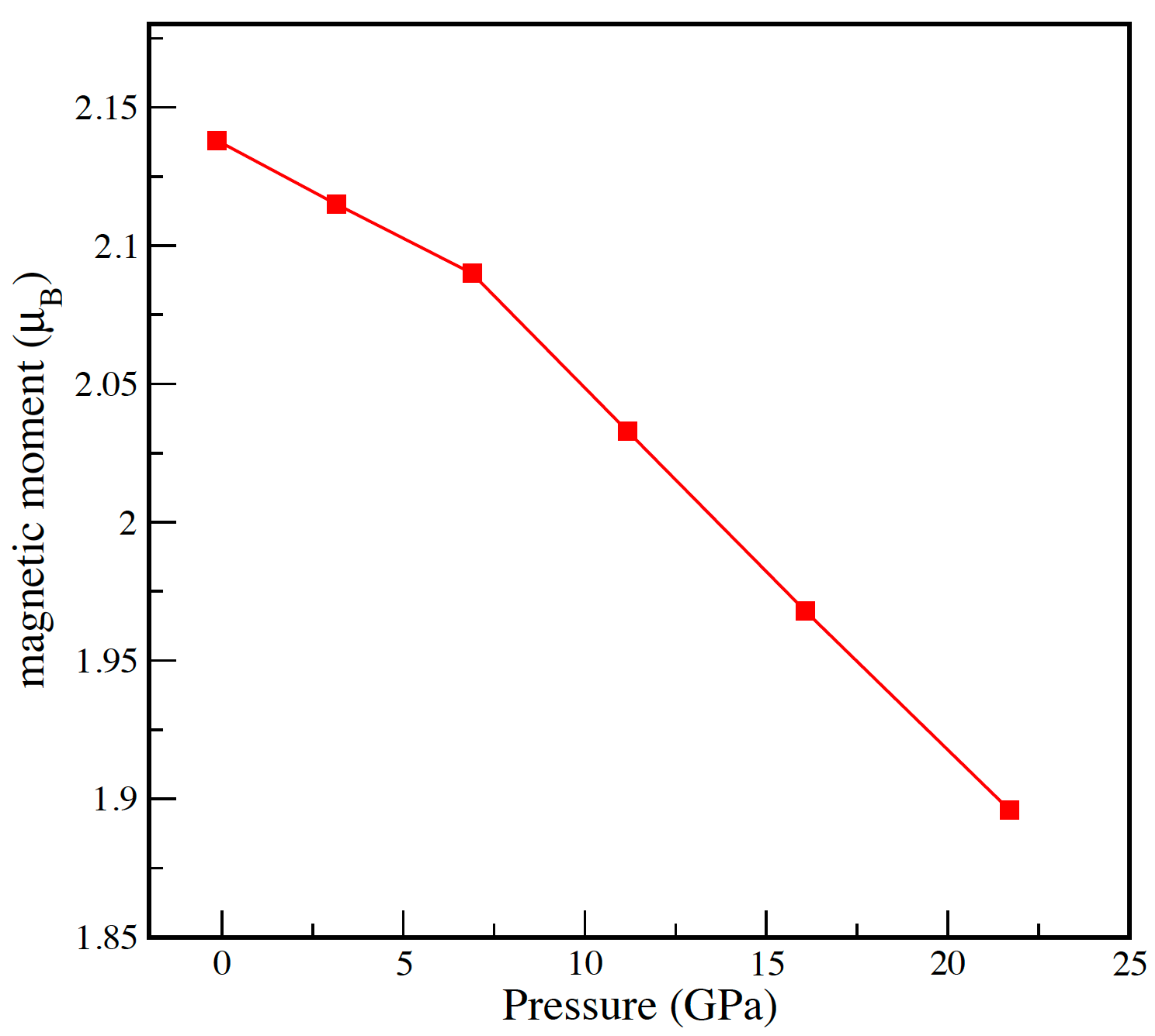}
    \caption{Pressure dependence of the calculated local Fe magnetic moment in $Fe_3Sn_2$, showing a gradual reduction of the moment amplitude under compression.}
    \label{moment}    
    \end{center}
\end{figure}

To observe the evolution of magnetic behavior, we first examined the pressure dependence of the local Fe magnetic moment (shown in Fig.~\ref{moment}). A gradual decrease in the moment, in line with the XES outcome, is observed with increasing pressure, indicating a weakening of the exchange splitting as Fe--Fe distances shrink and orbital hybridization strengthens. This trend reflects the suppression of the ferromagnetic component within the rhombohedral phase and provides the microscopic basis for understanding the subsequent SOC-driven anisotropic effects that emerge at higher pressures.

Next, the inter-atomic exchange interactions were evaluated for two representative structures: ambient condition and 21.7 GPa pressure ($-5\%$ volume change). The results are summarized in Table~\ref{exchange-coupling}. All three leading couplings ($J_1$, $J_2$, and $J_3$ (as indicated in Fig.~\ref{exchange}) are ferromagnetic and remain positive under compression. Upon applying hydrostatic pressure, the Fe-Fe distances decrease and the exchange interactions, particularly $J_2$, become stronger. This enhancement reflects the increased orbital overlap between neighboring Fe atoms.

\begin{table}[t]
\caption{Exchange interactions ($J_i$) between Fe-Fe ions in Fe$_3$Sn$_2$ at different hydrostatic strain values. Distances are in~\AA{} and exchange strengths in meV.}
\label{exchange-coupling}
\centering
\begin{tabular}{lccc}
\hline\hline
\textbf{$p$ (GPa)} & \textbf{Interaction} & \textbf{Distance (Å)} & \textbf{$J_i$ (meV)} \\
\hline
\multirow{3}{*}{0 (ambient)} 
  & $J_1$ & 2.48 & 16.18 \\
  & $J_2$ & 2.54 & 22.37 \\
  & $J_3$ & 2.80 & 23.61 \\
\hline
\multirow{3}{*}{21.7 ($-5\%$)} 
  & $J_1$ & 2.39 & 16.83 \\
  & $J_2$ & 2.52 & 25.88 \\
  & $J_3$ & 2.55 & 17.36 \\
\hline\hline
\end{tabular}
\end{table}
\par 
Following the strengthening of the isotropic exchange interactions, we next examined the evolution of the magnetocrystalline anisotropy (MAE) under pressure to assess the role of spin-orbit coupling (SOC). As summarized in Table~\ref{MAE}, Fe$_3$Sn$_2$ exhibits an in-plane easy axis at ambient conditions, with the $a$- and $b$-axis spin orientations lying lower in energy than the $c$-axis by 0.86~meV. Upon applying hydrostatic pressure (corresponding to 21.7~GPa), this energy difference increases to 1.32~meV, while the $a$- and $b$-axis energies remain degenerate. The larger magnitude of the MAE clearly indicates a strengthening of SOC-driven anisotropy under pressure, reinforcing the stabilization of an in-plane magnetic alignment relative to the out-of-plane ($c$-axis) direction. Such an enhancement of SOC with compression is further manifested in the behavior of the anisotropic exchange interactions. Our calculations suggest that the symmetric anisotropic exchange components $C_{ij}$s are very weak and we can ignore them in setting up spin Hamiltonian of this system. However,  as listed in Table~\ref{DM_interaction}, the antisymmetric Dzyaloshinskii-Moriya (DM) vectors $D_{ij}$ increase in magnitude for several Fe-Fe pairs, most notably for the second-nearest neighbors, where $|D_y|$ grows substantially from $0.42$ to $0.83$~meV, accompanied by an increase in $|D_z|$. The magnitude of $D_{ij}$ for second neighbor almost gets doubled due to the application of pressure of 21.7 GPa.  Since DM interaction originates exclusively from SOC, their growth under compression provides a microscopic description of the pressure-enhanced SOC strength inferred from MAE trends. The amplification of DM interactions, in particular, favors a canting of the Fe spins promoting noncollinear or weakly canted configurations within the kagome bilayers. This theoretical trend is consistent with our XMCD measurements, which reveal a progressive tilting of the ferromagnetic component away from the $c$-axis under pressure. Thus, the combined evolution of MAE and anisotropic exchange couplings under pressure establishes SOC-enhanced spin canting as a key precursor to the experimentally observed suppression of long-range ferromagnetic order. Taken together, these results indicate that pressure does not immediately suppress ferromagnetism through a collapse of isotropic exchange, but instead progressively reshapes the magnetic anisotropy landscape through enhanced spin--orbit coupling. This mechanism naturally explains why a tilted ferromagnetic configuration remains stable over a broad pressure range before the eventual disappearance of long-range magnetic order observed experimentally.

\begin{table}[t]
\caption{Comparison of magnetocrystalline anisotropy energy (MAE) values at ambient conditions and under applied pressure. The $c$-axis is taken as the reference axis.}
\label{MAE}
\centering
\setlength{\tabcolsep}{10pt} 
\begin{tabular}{lccc}
\hline\hline
\textbf{Pressure (GPa)} & \multicolumn{3}{c}{\textbf{Spin-axis Energy (meV)}} \\
\cline{2-4}
 & \textbf{$c$-axis} & \textbf{$b$-axis} & \textbf{$a$-axis} \\
\hline
Ambient (0) & 0.00 & $-0.86$ & $-0.86$ \\
21.7 ($-5\%$) & 0.00 & $-1.32$ & $-1.32$ \\
\hline\hline
\end{tabular}
\end{table}

\begin{table}[t]
\centering
\setlength{\tabcolsep}{0pt} 
\caption{Antisymmetric anisotropic Dzyaloshinskii–Moriya (DM) interaction strengths ($D$) at ambient conditions and under applied pressure. Values are in meV.}
\label{DM_interaction}
\begin{tabular}{ccccc}
\hline\hline
\textbf{Neighbor} & \textbf{0 GPa}    & \textbf{\textbar{}D$_{ij}$\textbar{}} & \textbf{21.7 GPa ($-5\%$)} &\textbf{\textbar{}D$_{ij}$\textbar{}} \\
\hline\\
1$^{st}$       & (0.14,0.08,0.14)   & 0.21                              & (0.16,0.09,0.02)    & 0.18                              \\\\
2$^{nd}$       & (0.00,-0.42,-0.25) & 0.48                              & (0.00,-0.83,-0.40)  & 0.92                              \\\\
3$^{rd}$       & (0.00,0.39,-0.09)  & 0.40                              & (0.00,0.11,-0.52)   & 0.53     \\
\hline
\end{tabular}
\end{table}

\begin{figure}[t]
    \begin{center}
     \includegraphics[width=1.0\columnwidth]{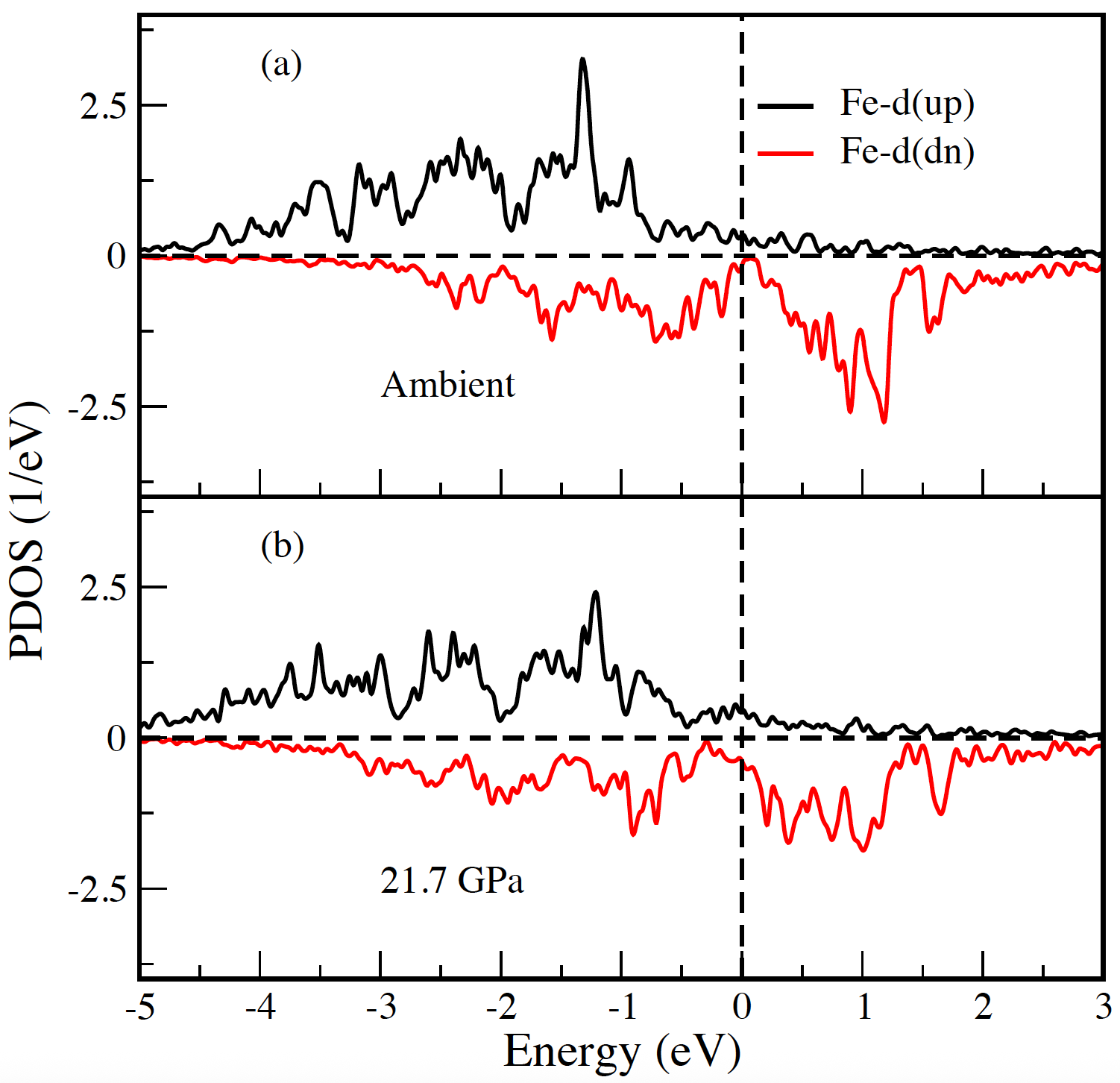}
    \caption {Partial density of states (PDOS) obtained from GGA calculations for Fe-3d orbital at (a) ambient (0 GPa) conditions and (b) at 21.7 GPa ($-5\%$).}
    \label{dos-band}    
    \end{center}
\end{figure}

\subsection{Electronic structure under application of pressure}
In addition, we analyzed the \textit{partial density of states} (PDOS) for the Fe~3$d$ orbitals under ambient and applied pressure $-5\%$ (21.7 GPa). In both the cases, finite spectral weight from Fe~3$d$ orbitals is observed at the Fermi level, confirming that Fe$_3$Sn$_2$ retains a metallic character over the entire studied pressure range. The overall bandwidth of the Fe~3$d$ states increases slightly under pressure, indicating enhanced orbital hybridization.

\section{New Conclusion}
In this work, we have experimentally investigated the interplay between crystal structure and magnetism in the kagome compound $Fe_3Sn_2$ under high pressure. Using a combination of X-ray diffraction, X-ray Magnetic Circular Dichroism, X-ray Emission Spectroscopy, and density functional theory calculations, we have shown that $Fe_3Sn_2$ undergoes a pressure-induced structural transition toward a lower-symmetry phase, which coincides with a drastic modification of its magnetic properties.

By combining X-ray Emission Spectroscopy and XMCD, we were able to disentangle the evolution of the amplitude of the local Fe magnetic moment from that of its orientation. XES reveals a clear high-spin to low-spin transition occurring around 20~GPa, concomitant with the structural transition identified by XRD and supported by evolutionary structural searches. This result demonstrates a strong coupling between lattice degrees of freedom and the local spin state, highlighting the sensitivity of the Fe $3d$ electrons to pressure-induced changes in bonding and bandwidth.

XMCD measurements further show that, while XES captures the collapse of the local moment amplitude, the magnetic orientation follows a distinct pressure–temperature evolution. At 300~K, the magnetic structure with moments aligned predominantly along the $c$ axis remains stable up to the structural transition around 20~GPa. At low temperature (30~K), a tilted magnetic configuration persists up to much higher pressures, reaching at least 33~GPa, indicating an enhanced robustness of the low-temperature magnetic phase against compression. At intermediate temperature (100~K), pressure stabilizes the low-temperature tilted phase already around 10~GPa, and this phase remains stable up to approximately 30~GPa. These results demonstrate that pressure destabilizes the high-temperature magnetic phase in favor of the low-temperature one, shifting the magnetic phase boundary to higher temperatures.

Spin-polarized DFT calculations provide a microscopic framework for understanding these experimental observations. While the isotropic ferromagnetic exchange interactions remain robust and even slightly enhanced under compression, the local Fe magnetic moment gradually decreases with pressure, in agreement with the XES results. More importantly, our calculations reveal a pronounced enhancement of spin–orbit–coupling-driven anisotropic interactions under pressure, manifested by an increase of the magnetocrystalline anisotropy energy and a substantial amplification of the Dzyaloshinskii–Moriya interaction. These effects favor a canting of the magnetic moments away from the $c$ axis and provide a natural explanation for the pressure-induced stabilization of the tilted magnetic phase observed by XMCD.
Taken together, our results establish pressure as a powerful tuning parameter that reshapes the magnetic energy landscape of $Fe_3Sn_2$ by simultaneously acting on structural symmetry, spin state, and magnetic anisotropy. The pressure-induced reduction of crystal symmetry and the concomitant enhancement of spin–orbit coupling destabilize the high-temperature collinear magnetic phase in favor of a canted configuration that becomes stable over an extended pressure–temperature region.

The fact that the high-pressure structure is only partially captured by DFT suggests that additional structural complexity, such as disorder, superstructural modulations, or subtle atomic rearrangements, may play a role in stabilizing the observed magnetic phases. More broadly, our findings highlight the central role of spin–lattice coupling in kagome magnets and demonstrate how external pressure can be used to selectively tune magnetic orientation independently from the local moment amplitude. Given the strong impact of spin orientation on the topological electronic structure of kagome systems, these results open new perspectives for pressure-driven control of topological and magnetic phases in $Fe_3Sn_2$ and related materials.

\section{Acknowledgments}
We acknowledge SOLEIL for providing the synchrotron beamtime on ODE (proposals 20201513, 20211491 and 99220160) PSICHE (20220361, 99220002) and GALAXIES beamlines and the MORPHEUS platform at the Laboratoire de Physique des Solides for sample orientation and alignment. K.M. acknowledges the funding support from the Department of Science and Technology, Government of India, through a core research grant (Grant No. CRG/2022/001826). V. B. acknowledges the Paris Ile-de-France Region in the framework of DIM MaTerRE (project DAC-VX).

\bibliography{biblio.bib}

@article{Wang2016,
  title = {Anomalous Hall effect in a ferromagnetic ${\mathrm{Fe}}_{3}{\mathrm{Sn}}_{2}$ single crystal with a geometrically frustrated Fe bilayer kagome lattice},
  author = {Wang, Qi and Sun, Shanshan and Zhang, Xiao and Pang, Fei and Lei, Hechang},
  journal = {Phys. Rev. B},
  volume = {94},
  issue = {7},
  pages = {075135},
  numpages = {5},
  year = {2016},
  month = {Aug},
  publisher = {American Physical Society},
  doi = {10.1103/PhysRevB.94.075135},
  url = {https://link.aps.org/doi/10.1103/PhysRevB.94.075135}
}

@article{Tao2023,
  title = {Investigating the magnetoelastic properties in FeSn and ${\mathrm{Fe}}_{3}{\mathrm{Sn}}_{2}$ flat band metals},
  author = {Tao, Yu and Daemen, Luke and Cheng, Yongqiang and Neuefeind, Joerg C. and Louca, Despina},
  journal = {Phys. Rev. B},
  volume = {107},
  issue = {17},
  pages = {174407},
  numpages = {9},
  year = {2023},
  month = {May},
  publisher = {American Physical Society},
  doi = {10.1103/PhysRevB.107.174407},
  url = {https://link.aps.org/doi/10.1103/PhysRevB.107.174407}
}

@article{Sales2019,
  title = {Electronic, magnetic, and thermodynamic properties of the kagome layer compound FeSn},
  author = {Sales, Brian C. and Yan, Jiaqiang and Meier, William R. and Christianson, Andrew D. and Okamoto, Satoshi and McGuire, Michael A.},
  journal = {Phys. Rev. Mater.},
  volume = {3},
  issue = {11},
  pages = {114203},
  numpages = {9},
  year = {2019},
  month = {Nov},
  publisher = {American Physical Society},
  doi = {10.1103/PhysRevMaterials.3.114203},
  url = {https://link.aps.org/doi/10.1103/PhysRevMaterials.3.114203}
}

@article{Liu2020,
  title = {Strain-induced pseudomagnetic field and quantum oscillations in kagome crystals},
  author = {Liu, Tianyu},
  journal = {Phys. Rev. B},
  volume = {102},
  issue = {4},
  pages = {045151},
  numpages = {8},
  year = {2020},
  month = {Jul},
  publisher = {American Physical Society},
  doi = {10.1103/PhysRevB.102.045151},
  url = {https://link.aps.org/doi/10.1103/PhysRevB.102.045151}
}

@article{Ren2022,
archivePrefix = {arXiv},
arxivId = {2202.04177},
author = {Ren, Zheng and Li, Hong and Sharma, Shrinkhala and Bhattarai, Dipak and Zhao, He and Rachmilowitz, Bryan and Bahrami, Faranak and Tafti, Fazel and Fang, Shiang and Ghimire, Madhav Prasad and Wang, Ziqiang and Zeljkovic, Ilija},
doi = {10.1038/s41535-022-00521-y},
eprint = {2202.04177},
isbn = {4153502200},
issn = {23974648},
journal = {npj Quantum Materials},
number = {1},
pages = {1--7},
publisher = {Springer US},
title = {{Plethora of tunable Weyl fermions in kagome magnet Fe3Sn2 thin films}},
volume = {7},
year = {2022}
}

@article{Kang2020,
archivePrefix = {arXiv},
arxivId = {1906.02167},
author = {Kang, Mingu and Ye, Linda and Fang, Shiang and You, Jhih Shih and Levitan, Abe and Han, Minyong and Facio, Jorge I. and Jozwiak, Chris and Bostwick, Aaron and Rotenberg, Eli and Chan, Mun K. and McDonald, Ross D. and Graf, David and Kaznatcheev, Konstantine and Vescovo, Elio and Bell, David C. and Kaxiras, Efthimios and van den Brink, Jeroen and Richter, Manuel and {Prasad Ghimire}, Madhav and Checkelsky, Joseph G. and Comin, Riccardo},
doi = {10.1038/s41563-019-0531-0},
eprint = {1906.02167},
issn = {14764660},
journal = {Nature Materials},
number = {2},
pages = {163--169},
pmid = {31819211},
publisher = {Springer US},
title = {{Dirac fermions and flat bands in the ideal kagome metal FeSn}},
url = {http://dx.doi.org/10.1038/s41563-019-0531-0},
volume = {19},
year = {2020}
}

@article{Malaman1978,
author = {Malaman, B and Fruchart, D and Caer, G Le},
doi = {10.1088/0305-4608/8/11/022},
issn = {0305-4608},
journal = {Journal of Physics F: Metal Physics},
month = {nov},
number = {11},
pages = {2389--2399},
title = {{Magnetic properties of Fe 3 Sn 2 . II. Neutron diffraction study (and Mossbauer effect)}},
url = {https://iopscience.iop.org/article/10.1088/0305-4608/8/11/022},
volume = {8},
year = {1978}
}

@article{Baudelet2016,
author = {Baudelet, F. and Nataf, L. and Torchio, R.},
doi = {10.1080/08957959.2016.1211274},
issn = {14772299},
journal = {High Pressure Research},
number = {3},
pages = {429--444},
title = {{New scientific opportunities for high pressure research by energy-dispersive XMCD}},
volume = {36},
year = {2016}
}

@article{Baudelet2011,
author = {Baudelet, F. and Kong, Q. and Nataf, L. and Cafun, J. D. and Congeduti, A. and Monza, A. and Chagnot, S. and Iti{\'{e}}, J. P.},
doi = {10.1080/08957959.2010.532794},
issn = {08957959},
journal = {High Pressure Research},
number = {1},
pages = {136--139},
title = {{ODE: A new beam line for high-pressure XAS and XMCD studies at SOLEIL}},
volume = {31},
year = {2011}
}

@article{Fenner2009,
archivePrefix = {arXiv},
arxivId = {1809.05466},
author = {Fenner, L. A. and Dee, A. A. and Wills, A. S.},
doi = {10.1088/0953-8984/21/45/452202},
eprint = {1809.05466},
issn = {09538984},
journal = {Journal of Physics Condensed Matter},
number = {45},
title = {{Non-collinearity and spin frustration in the itinerant kagome ferromagnet Fe3Sn2}},
volume = {21},
year = {2009}
}

@article{Shen2020,
author = {Shen, Guoyin and Wang, Yanbin and Dewaele, Agnes and Wu, Christine and Fratanduono, Dayne E. and Eggert, Jon and Klotz, Stefan and Dziubek, Kamil F. and Loubeyre, Paul and Fat'yanov, Oleg V. and Asimow, Paul D. and Mashimo, Tsutomu and Wentzcovitch, Renata M.M.},
doi = {10.1080/08957959.2020.1791107},
issn = {14772299},
journal = {High Pressure Research},
pages = {299--314},
title = {{Toward an international practical pressure scale: A proposal for an IPPS ruby gauge (IPPS-Ruby2020)}},
year = {2020}
}

@article{Tateiwa2009,
archivePrefix = {arXiv},
arxivId = {1001.0454},
author = {Tateiwa, Naoyuki and Haga, Yoshinori},
doi = {10.1063/1.3265992},
eprint = {1001.0454},
isbn = {0034-6748 VO - 80},
issn = {00346748},
journal = {Review of Scientific Instruments},
number = {12},
pmid = {20059148},
title = {{Evaluations of pressure-transmitting media for cryogenic experiments with diamond anvil cell}},
volume = {80},
year = {2009}
}

@article{Boettger2012,
author = {Boettger, Jonathan C. and Honnell, Kevin G. and Peterson, Jeffrey H. and Greeff, Carl W. and Crockett, Scott D.},
doi = {10.1063/1.3686402},
isbn = {9780735410060},
issn = {0094243X},
journal = {AIP Conference Proceedings},
number = {June 2014},
pages = {812--815},
title = {{Tabular equation of state for gold}},
volume = {1426},
year = {2012}
}

@article{Giefers2006,
author = {Giefers, Hubertus and Nicol, Malcolm},
doi = {10.1016/j.jallcom.2005.11.061},
issn = {09258388},
journal = {Journal of Alloys and Compounds},
number = {1-2},
pages = {132--144},
title = {{High pressure X-ray diffraction study of all Fe-Sn intermetallic compounds and one Fe-Sn solid solution}},
volume = {422},
year = {2006}
}

@article{Birch1947,
author = {Birch, Francis},
doi = {10.1103/PhysRev.71.809},
issn = {0031899X},
journal = {Physical Review},
number = {11},
pages = {809--824},
title = {{Finite elastic strain of cubic crystals}},
volume = {71},
year = {1947}
}

@article{Rueff2015,
author = {Rueff, J. P. and Ablett, J. M. and C{\'{e}}olin, D. and Prieur, D and Moreno, Th and Bal{\'{e}}dent, V. and Lassalle-Kaiser, B. and Rault, J. E. and Simon, M. and Shukla, A},
doi = {10.1107/S160057751402102X},
issn = {16005775},
journal = {Journal of Synchrotron Radiation},
number = {1},
pages = {175--179},
pmid = {25537606},
title = {{The GALAXIES beamline at the SOLEIL synchrotron: Inelastic X-ray scattering and photoelectron spectroscopy in the hard X-ray range}},
volume = {22},
year = {2015}
}

@article{Oganov2006,
    author = {Oganov, Artem R. and Glass, Colin W.},
    title = {Crystal structure prediction using ab initio evolutionary techniques: Principles and applications},
    journal = {The Journal of Chemical Physics},
    volume = {124},
    number = {24},
    pages = {244704},
    year = {2006},
    month = {06},
    issn = {0021-9606},
    doi = {10.1063/1.2210932},
    url = {https://doi.org/10.1063/1.2210932},
    eprint = {https://pubs.aip.org/aip/jcp/article-pdf/doi/10.1063/1.2210932/13690624/244704\_1\_online.pdf},
}

@article{Oganov2011,
author = {Oganov, Artem R. and Lyakhov, Andriy O. and Valle, Mario},
doi = {10.1021/AR1001318/ASSET/IMAGES/MEDIUM/AR-2010-001318_0001.GIF},
issn = {00014842},
journal = {Accounts of Chemical Research},
month = {mar},
number = {3},
pages = {227--237},
publisher = {American Chemical Society},
title = {{How evolutionary crystal structure prediction works-and why}},
url = {/doi/pdf/10.1021/ar1001318},
volume = {44},
year = {2011}
}

@article{Lyakhov2013,
title = {New developments in evolutionary structure prediction algorithm USPEX},
journal = {Computer Physics Communications},
volume = {184},
number = {4},
pages = {1172-1182},
year = {2013},
issn = {0010-4655},
doi = {https://doi.org/10.1016/j.cpc.2012.12.009},
url = {https://www.sciencedirect.com/science/article/pii/S0010465512004055},
author = {Andriy O. Lyakhov and Artem R. Oganov and Harold T. Stokes and Qiang Zhu},
}

@article{Ong2024,
author = {Ong, Chin Shen and Donzel-Gargand, Olivier and Berastegui, Pedro and Cedervall, Johan and {Bayrak Pehlivan}, Ilknur and Hervoches, Charles and Beran, Premysl and Edvinsson, Tomas and Eriksson, Olle and Jansson, Ulf},
doi = {10.1021/acs.inorgchem.4c00560},
issn = {1520510X},
journal = {Inorganic Chemistry},
number = {23},
pages = {10490--10499},
pmid = {38801717},
title = {{The Crystal Structure of Al4SiC4 Revisited}},
url = {https://doi.org/10.1021/acs.inorgchem.4c00560},
volume = {63},
year = {2024}
}

@article{Kresse1999,
  title = {From ultrasoft pseudopotentials to the projector augmented-wave method},
  author = {Kresse, G. and Joubert, D.},
  journal = {Phys. Rev. B},
  volume = {59},
  issue = {3},
  pages = {1758--1775},
  numpages = {0},
  year = {1999},
  month = {Jan},
  publisher = {American Physical Society},
  doi = {10.1103/PhysRevB.59.1758},
  url = {https://link.aps.org/doi/10.1103/PhysRevB.59.1758}
}

@article{Zhipeng2017,
author = {Hou, Zhipeng and Ren, Weijun and Ding, Bei and Xu, Guizhou and Wang, Yue and Yang, Bing and Zhang, Qiang and Zhang, Ying and Liu, Enke and Xu, Feng and Wang, Wenhong and Wu, Guangheng and Zhang, Xixiang and Shen, Baogen and Zhang, Zhidong},
title = {Observation of Various and Spontaneous Magnetic Skyrmionic Bubbles at Room Temperature in a Frustrated Kagome Magnet with Uniaxial Magnetic Anisotropy},
journal = {Advanced Materials},
volume = {29},
number = {29},
pages = {1701144},
doi = {https://doi.org/10.1002/adma.201701144},
year = {2017}
}

@article{Balents2010,
  author    = {Balents, Leon},
  title     = {Spin liquids in frustrated magnets},
  journal   = {Nature},
  volume    = {464},
  pages     = {199--208},
  year      = {2010},
  doi       = {10.1038/nature08917}
}

@article{Zhou2017,
  author    = {Zhou, Yi and Kanoda, Kazushi and Ng, Tai-Kai},
  title     = {Quantum spin liquid states},
  journal   = {Reviews of Modern Physics},
  volume    = {89},
  pages     = {025003},
  year      = {2017},
  doi       = {10.1103/RevModPhys.89.025003}
}

@article{Ye2018,
  author    = {Ye, Linda and Kang, Mingu and Liu, Junwei and...},
  title     = {Massive Dirac fermions in a ferromagnetic kagome metal},
  journal   = {Nature},
  volume    = {555},
  pages     = {638--642},
  year      = {2018},
  doi       = {10.1038/nature25987}
}

@article{Yin2019,
  author    = {Yin, Jia-Xin and Zhang, Songtian S. Zhang and...},
  title     = {Negative flat band magnetism in a spin--orbit-coupled correlated kagome magnet},
  journal   = {Nature Physics},
  volume    = {15},
  pages     = {443--448},
  year      = {2019},
  doi       = {10.1038/s41567-018-0410-7}
}

@article{Liu2018,
  author    = {Liu, Enke and Sun, Yan and Kumar, Nitesh and...},
  title     = {Giant anomalous Hall effect in a ferromagnetic kagome-lattice semimetal},
  journal   = {Nature Physics},
  volume    = {14},
  pages     = {1125--1131},
  year      = {2018},
  doi       = {10.1038/s41567-018-0234-5}
}

@article{Wang2012,
  author    = {Wang, Lin and Mao, Ho-Kwang and Hemley, Russell J.},
  title     = {Coupling between structural and magnetic transitions under pressure},
  journal   = {Reports on Progress in Physics},
  volume    = {75},
  pages     = {066501},
  year      = {2012},
  doi       = {10.1088/0034-4885/75/6/066501}
}

@article{Kunes2010,
  author    = {Kune{\v s}, Jan and Lukoyanov, Alexey V. and Anisimov, Vladimir I. and...},
  title     = {Pressure-driven spin-state transition in correlated materials},
  journal   = {Physical Review B},
  volume    = {81},
  pages     = {035122},
  year      = {2010},
  doi       = {10.1103/PhysRevB.81.035122}
}

@article{Lou2024,
  title = {Orbital-selective effect of spin reorientation on the Dirac fermions in a non-charge-ordered kagome ferromagnet Fe$_3$Ge},
  author = {Lou, Rui and Zhou, Liqin and Song, Wenhua and Fedorov, Alexander and Tu, Zhijun and Jiang, Bei and Wang, Qi and Li, Man and Liu, Zhonghao and Chen, Xuezhi and Rader, Oliver and B{\"u}chner, Bernd and Sun, Yujie and Weng, Hongming and Lei, Hechang and Wang, Shancai},
  journal = {Nature Communications},
  volume = {15},
  pages = {9823},
  year = {2024},
  publisher = {Nature Publishing Group},
  doi = {10.1038/s41467-024-51243-5}
}

@article{vasp1,
  title="{Ab initio molecular dynamics for liquid metals}",
  author = {Kresse, G. and Hafner, J.},
  journal = {Phys. Rev. B},
  volume = {47},
  issue = {1},
  pages = {558--561},
  numpages = {0},
  year = {1993},
  month = {Jan},
  publisher = {American Physical Society},
  doi = {10.1103/PhysRevB.47.558},
  url = {https://link.aps.org/doi/10.1103/PhysRevB.47.558}
}

@article{vasp2,
  title="{Efficient iterative schemes for ab initio total-energy calculations using a plane-wave basis set}",
  author = {Kresse, G. and Furthm\"uller, J.},
  journal = {Phys. Rev. B},
  volume = {54},
  issue = {16},
  pages = {11169--11186},
  numpages = {0},
  year = {1996},
  month = {Oct},
  publisher = {American Physical Society},
  doi = {10.1103/PhysRevB.54.11169},
  url = {https://link.aps.org/doi/10.1103/PhysRevB.54.11169}
}

@article{RSPt,
  title="{Exchange parameters of strongly correlated materials: Extraction from spin-polarized density functional theory plus dynamical mean-field theory}",
  author = {Kvashnin, Y. O. and Gr\aa{}n\"as, O. and Di Marco, I. and Katsnelson, M. I. and Lichtenstein, A. I. and Eriksson, O.},
  journal = {Phys. Rev. B},
  volume = {91},
  issue = {12},
  pages = {125133},
  numpages = {10},
  year = {2015},
  month = {Mar},
  publisher = {American Physical Society},
  doi = {10.1103/PhysRevB.91.125133},
  url = {https://link.aps.org/doi/10.1103/PhysRevB.91.125133}
}

@article{Birch,
  title = {Finite Elastic Strain of Cubic Crystals},
  author = {Birch, Francis},
  journal = {Phys. Rev.},
  volume = {71},
  issue = {11},
  pages = {809--824},
  numpages = {0},
  year = {1947},
  month = {Jun},
  publisher = {American Physical Society},
  doi = {10.1103/PhysRev.71.809},
  url = {https://link.aps.org/doi/10.1103/PhysRev.71.809}
}

@article{murnaghan,
  title={The compressibility of media under extreme pressures},
  author={Murnaghan, Francis Dominic},
  journal={Proceedings of the National Academy of Sciences},
  volume={30},
  number={9},
  pages={244--247},
  year={1944}
}

@article{kargeti2025charge,
  title={Charge-state dependent spin-orbit coupling and quantum phase transitions in Ir-Ru oxides},
  author={Kargeti, Kuldeep and Mallick, Bidyut and Borisov, Vladislav and Ali, Sk Soyeb and Hellsvik, Johan and Eriksson, Olle and Panda, SK},
  journal={Physical Review B},
  volume={111},
  number={19},
  pages={195148},
  year={2025},
  publisher={APS}
}

@article{kargeti2024strain,
  title={Strain-induced electronic and magnetic transition in the S= 3 2 antiferromagnetic spin chain compound LaCrS 3},
  author={Kargeti, Kuldeep and Sen, Aadit and Panda, SK},
  journal={Physical Review B},
  volume={109},
  number={3},
  pages={035125},
  year={2024},
  publisher={APS}
}

@article{Li2022,
  title        = {Spin-polarized imaging of the antiferromagnetic structure and field-tunable bound states in kagome magnet FeSn},
  author       = {Li, H. and Zhao, H. and Yin, Q. and et al.},
  journal      = {Scientific Reports},
  volume       = {12},
  pages        = {14525},
  year         = {2022},
  doi          = {10.1038/s41598-022-18678-8},
  url          = {https://doi.org/10.1038/s41598-022-18678-8}
}

@article{Ablett2025,
  title        = {MULTIXS: A new scanning multi-analyzer x-ray emission spectrometer at the GALAXIES beamline at synchrotron SOLEIL},
  author       = {Ablett, J. M. and Berlioux, A. and Prieur, D. and Harrison, J. and Heller, L. and Gliga, S. and Rueff, J.-P.},
  journal      = {Review of Scientific Instruments},
  volume       = {96},
  number       = {5},
  pages        = {053104},
  year         = {2025},
  doi          = {10.1063/5.0250429},
  url          = {https://doi.org/10.1063/5.0250429}
}
\end{document}